\documentclass[manuscript,aps]{revtex4}
\usepackage{graphicx}% Include figure files
\usepackage{dcolumn}% Align table columns on decimal point
\usepackage{bm}% bold math

\begin{document}

\preprint{}

\title{Transverse Dynamics and Energy Tuning of Fast Electrons Generated in
Sub-Relativistic Intensity Laser Pulse Interaction with Plasmas}% Force line breaks with \\

\author{Michiaki Mori\footnote{E-mail: mori.michiaki@jaea.go.jp}}
\author{Masaki Kando}%
\author{Izuru Daito}%
\author{Hideyuki Kotaki}%
\author{Yukio Hayashi}%
\author{Atsushi Yamazaki}%
\author{Koichi Ogura}%
\author{Akito Sagisaka}%
\author{James Koga}%
\author{Kazuhisa Nakajima}%
\author{Hiroyuki Daido}%
\author{Sergei V. Bulanov}%
\author{Toyoaki Kimura}%

\address{Advanced Photon Research Center, Japan Atomic Energy Agency,\\ Umemidai 8-1, Kizu, Kyoto, 619-0215 Japan}

\date{\today}% It is always \today, today,
             %  but any date may be explicitly specified

\begin{abstract}
The regimes of quasi-mono-energetic electron beam generation were experimentally studied in
the sub-relativistic intensity laser plasma interaction. The observed electron acceleration regime is unfolded with two-dimensional-particle-in-cell simulations of laser-wakefield generation in the self-modulation regime.
\end{abstract}

% PACS, the Physics and Astronomy
                             % Classification Scheme.
%\keywords{Suggested keywords}%Use showkeys class option if keyword
                              %display desired
\maketitle
\section{Introduction}

Laser accelerator \cite{T-D} development is now entering a new mature stage
in which it has become possible to manipulate in a controllable way the
parameters of accelerated charged particle beams. For a broad range of
applications, high stability in ultrarelativistic electron production is
required, which implies high directionality, controllability and
reproducibility in the accelerated electron energy and collimation.
Quasi-monoenergetic electron beam generation has been observed in several
laser-plasma interaction experiments, the results of which are presented in
Refs. \cite{mono1,mono2}. This breakthrough resulted from many previous
investigations of laser-driven charged particle acceleration conducted by a
number of scientific groups (see Ref. \cite{LA}), where the high energy
electron production and huge accelerating gradients ($\approx $ 100 GV/m)
have been observed. 

In a majority of the above mentioned experiments, the
intensity of the laser light at the focus was relativistic, i.e. the
dimensionless laser field amplitude, $a_0 = e E_0/m_e \omega c$, was greater
than unity: $a_{0} \approx 0.85 [(I/10^{18}\ W/cm^2) (\lambda/1 \mu
m)^2]^{1/2}>1$. Here, $e$ is the electron charge, $E_0$ is the laser
pulse peak electric field, $m_e$ is the electron mass, $\omega$ is the laser
frequency, and $c$ is the speed of light in vacuum. Moreover, the plasma
density was also substantially large, e.g. the electron density at which the
quasi-monoenergetic electron production was observed \cite{mono1,mono2},
ranged from $6\times 10^{18}$ cm$^{-3}$ to $1.3\times 10^{20}$ cm$^{-3}$.
High enough electron density provides favorable conditions for self-focusing
and stimulated Raman scattering development resulting in laser pulse
self-modulation and the enhancement of the wake-field amplitude. This leads
to the wakewave-breaking and the electron injection into the acceleration
phase. On the other hand the nonlinear process development raises additional
questions on the directionality of the accelerated electron bunches, which
demonstrates a need of further thorough experimental studies of the
quasi-monoenergetic electron beam generation. 

Below, in this letter, we
present the the results of experiments on the electron acceleration during a
multi-terawatt femtosecond range laser interaction with an underdense plasma
under the conditions, when the self-modulated wakefield regime occurs 
\cite{S-M}. The acceleration regime is corroborated with 2D-particle-in-cell
computer simulations of the laser pulse interaction with the underdense
plasma.

\section{Experimental setup and condition}

The experiments are performed with a terawatt 10 Hz Ti:Sapphire CPA laser
system "JLITE-X" ($\lambda$=800 nm) at the Advanced Photon Research Center
of the Japan Atomic Energy Agency (JAEA-APRC). The laser pulse with 230 mJ
pulse energy on target has a typical duration (half maximum ) equal to 70
fs. The estimated peak power of the laser beam is 3 TW. The ratio of the
prepulse to the main pulse intensity is less than $10^{-5}$. The laser beam
is focused onto the target by using a $15 ^o$, 646 mm focal length, f/13
off-axis parabolic mirror. The spot size at the focus is $15 \mu m$ at
half maximum and $25 \mu m$ at $1/e^2$. The energy concentration is 
55 $ \% $ within the $1/e^2$ spot. The focal peak intensity is estimated to be 
$9 \times 10^{17} W/cm^2$. It corresponds to a dimensionless amplitude of
the laser field, $a_0$, equal to $0.67$. By evacuating the laser beam path
between the pulse compressor and the interaction chamber the pressure is
made to be less than $10^{-1}$ Pa. We measured the neutral gas
density in the target, a $600 \ \mu m$ long supersonic He gas-jet, by using 
a Mach-Zehnder interferometer.
The plasma density is calculated by doubling the neutral gas density under
the assumption of full helium ionization. This assumption is based on the
fact that for our experimental conditions the barrier-suppression-ionization
threshold intensity for He$^{2+}$ is equal to $3.9 \times 10^{16} W/cm^2$ 
\cite{AUG} and the tunneling ionization threshold is $8.8 \times 10^{15}
W/cm^2$. These thresholds are well below the laser intensity at focus, 
$\approx 10^{18} W/cm^2$. As a result the plasma density at the focus is
calculated to be $n_0=5\times 10^{19}$ cm$^{-3}$. For the above parameters
the wake wave wavelength, $\lambda_p = 2 \pi c /\omega_{pe} \le 10\lambda$,
is shorter than the pulse length, which is about $\approx 30 \lambda$. The
laser power, 3 TW, is above the threshold for the relativistic
self-focusing, $P>P_{cr}=17 (\omega/\omega_{pe})^2 GW$, \cite{SF}. 
Here $\omega_{pe}=(4 \pi n_0 e^2/m_e)^{1/2}$ is the Langmuir
frequency and $\lambda$ is the laser wavelength.

In order to detect an energetic electron spectrum, we use an electron
spectrometer (ESM) which is composed of a slit shape lead collimator (the
slit width, $w$, is equal to 5 mm), an electromagnet, and two-layered
Imaging Plates (IP) [BAS-SR, Fuji film product]. The spectrometer is placed
at the rear side of the target behind the interaction point. Each IP is
covered with 12 $\mu m$ thick Al foil in order to avoid exposure to visible
light and the low energy electron component. We note that the IP is
sensitive not only to the signal created by the fast electrons but also to
the X-rays. As a consequence of this, there is a problem of separation of
the signal produced by the electrons from the noise originating from the
X-rays. In order to obtain an IP image which is predominantly formed by the
electron beam and to ascertain whether the electron beam contains a
quasi-monoenergetic component, we varied the magnetic field value from 0 to
0.13 T by changing the drive electric current in the electromagnet. As a
result, we found that X-ray signal effects are not so important for
estimating the electron number.

\section{Experimental results}

Figure 1 shows the typical energy resolved electron beam image obtained with
the ESM and the electron energy spectrum. Here, the horizontal coordinate
presents the energy profile at the IP and the vertical coordinate shows the
electron spatial distribution. The image is taken from a single shot. In
this case, the plasma density is set to be $4.7 \times 10^{19}$ cm$^{-3}$. A
well localized in space and energy spot is seen in the vicinity of the
energy equal to 19.6 MeV. The image spatial size at the IP in the vertical
direction is 5.7 mm at the half maximum. For the distance between the gas
jet and the IP equal to 760 mm, we find the electron beam divergence to be
about $7.5$ mrad. We point out that for lower plasma density, $n=3.1 \times
10^{19}$ cm$^{-3}$, the electron energy distribution taken with ten
accumulated shots, does not show a high energy component in the electron
beam. In order to obtain the electron number in the quasi-monoenergetic
components, we integrate the electron distribution at the IP over the
energy. By using the sensitivity of the IP (see Ref. \cite{ESM}), the total
charge of the quasi-monoenergetic electron component in the energy range
from $16.3$ to $25$ MeV and in the angle 5 mrad is found to be $\approx$0.8
pC with the energy spread approximately equal to 4.8 MeV at the half maximum.

We have also observed spatial-spectral split images. The typical spatial
(along the vertical axis) and spectral (along the horizontal axis) resolved
images are shown in Fig. 2. These images are taken in a single shot under
the same conditions as in the case presented in Fig. 1. In Fig. 2(a), we see
a single peak in space and a double peak in the energy distribution. The
energies of the double peak component are 13.5 MeV and 18.1 MeV,
respectively. In Fig. 2(b), a triple spatial split peak with different
quasi-monoenergetic energies is presented. Here, the peak energies of the
triple peak are 3.3 MeV, 4.5 MeV, and 7.5 MeV, respectively. The
directionality of electron beam with respect to the laser propagation
direction are -13 mrad, 6 mrad, and 20 mrad.

We are able to control the electron beam parameters by changing the plasma
density in the target. Figure 3 a) shows the electron energy spectra (each
of them is taken in a single shot) at four different values of the plasma
density. By increasing slightly the backing pressure of the gas jet the
plasma density is set to be equal to $4.1 \times 10^{19}$ cm$^{-3}$ (1), 
$4.7 \times 10^{19}$ cm$^{-3}$ (2), $5.0 \times 10^{19}$ cm$^{-3}$ (3), and 
$6.6 \times 10^{19}$ cm$^{-3}$ (4). This results in changing the position of
the quasi-monoenergetic peak from 8.5 MeV to 19.6 MeV. At the plasma density
above $4.7 \times10^{19}$ cm$^{-3}$, the energy width of the electron beam
becomes broadened.

\section{Results of 2D-PIC simulations}

In order to elucidate the acceleration regime responsible for the
quasi-monoenergetic electron beam formation, extensive two-dimensional PIC
simulations of the laser pulse interaction with the underdense plasma target
have been performed with the use of the REMP code \cite{REMP}. In the
simulations presented here a linearly p-polarized laser pulse with the
dimensionless amplitude $a_0 = 0.67$, corresponding to the peak intensity I
= 9 $\times 10^{17}$ W/cm$^{2}$ for $\lambda = 0.8 \ \mu m$ laser,
propagates along the $\mathit{x}$ axis. The laser pulse has a Gaussian
envelope with FWHM size 27 $\times$ 31 $\lambda ^2$. The plasma density is 
$n_0 = 2.32 \times 10^{-2} \ n_{cr}$, i.e. $n_0 =5 \times 10^{19}$ cm$^{-3}$,
which corresponds to $\omega_{pe}/\omega =0.153$. Ions have an absolute
charge two times larger than the electrons, and mass ratio $m_i = m_e \times
4 \times 1836$. The simulation box has 4500$\times$620 grid points with a 
0.1 $\lambda$ mesh size. The target has the form of an underdense plasma slab of
size 425$\times$60$\lambda^2$. The total number of quasiparticles is about 
$10^7$. The boundary conditions are absorbing in all directions for both the
electromagnetic (EM) radiation and the quasiparticles. The space and time
units are the wavelength $\lambda$ and the period $2\pi/\omega$ of the
incident radiation.

Under the conditions of simulations as in the case of the above discussed
experiment the laser pulse length is approximately 4 times larger than the
wake wavelength, $2 \pi c / \omega_{pe}$, and the observed laser pulse
evolution corresponds to the development of the self-modulation regime.
According to Ref. \cite{S-M} in this regime the modulational instability
results in the laser pulse modulations in the longitudinal direction and in
the generation of the wake plasma wake in and behind the laser pulse. In
Fig. 4 we present the electron density distribution in the $(x,y)$ plane at 
$t=300$, which shows a regular structure in the wake plasma wave left behind
the laser pulse. The wake wave has a finite length, which corresponds to the
finite time required for the modulational instability development. In frame
b) we plot the $E_z$ component of the electric field along the laser pulse
axis, $y=0$, for $t=300$. This clearly demonstrates that the initially
Gaussian laser pulse becomes modulated with a modulation wavelength equal to
the wake wave wavelength and its maximal amplitude becomes substantially
higher, by a factor larger than 2, than the initial pulse amplitude. In
frame c) we present the longitudinal, $E_x$, component of the electric field
along the laser pulse axis, $y=0$, for $t=300$, which characterizes the
evolution of the wake wave structure. We point out that the wake field
has a regular structure behind the laser pulse over a distance substantially
longer than the laser pulse length (see Fig. 4 c)).

The dependence of the electron energy spectrum on the target density found
from the PIC simulations (see Fig. 3 b)) has a behavior similar to that in
the experiment, which is seen in Fig. 3 (a). This clearly demonstrates the
typicalness of the electron acceleration regime realized in the present work.

In Fig. 5, projections of the electron phase space, a) plane $(x,p_x)$, b)
plane $(y,p_x)$, and c) plane $(x,p_y)$, at $t=400$ are plotted. The
transverse wake wave breaking of multiple wake cycles results in the
electron injection into the acceleration phase with the production of
multiple fast electron bunches seen in Fig. 5 a). At the time about $t=400$
the electrons reach their maximal energy of the order of $30$ MeV. The
electrons with the maximal energy are accelerated in the 2-nd wake period,
as seen in the electron phase plane, $(x,p_x)$, presented in Fig. 5 a). The
electron injection occurs not in the first cycle of the wake wave according
to the theory of the transverse wake wave breaking, which predicts that in a
tenuous plasma the wake wave breaks at a finite distance behind the wake
wave front, \cite{TWB}. In Figs. 5 a) and c) we indicate by the dashed line 
the position of the laser pulse leading edge. It is at $x=375$ and first electron 
bunch is localized at $x=365$. The electron energy spectrum has the form of a
quasi-monoenergetic bunch with a maximum at $20$ MeV (see Fig. 3 b), and
width $\approx 20\%$. The experimentally obtained form of the
quasi-monoenergetic electron beam as seen in Fig. 1 is in good agreement
with these results. Phase planes b) $(y,p_x)$ and c) $(x,p_y)$, show that
spatially separated bunches are accelerated in subsequent wake wave periods.
In the frames a) and b) the arrows A and B indicate the spatially separated
electron beamlets, which also have different energies. In the $(x,p_y)$
phase plane c), arrows indicate the aside beamlets. Inside the 2-nd wake
period, a quasi-monoenergetic electron bunch with its energy of about 20 MeV
is generated (indicated by the arrow A). 

In addition, broadened
quasi-monoenergetic electron bunches with energy approximately equal to 10
MeV can be seen in the 2-nd and 3-rd wake periods (indicated by the arrow
B). These features are in sufficiently good agreement with the multiple
quasi-mono energetic electron beam production observed in our experiments
and shown in Fig.2 (a). In addition, 2D-PIC simulations also demonstrate the
bending of electron beam bunches. The transverse electron momentum versus
the x-coordinate is shown in Fig. 5 c). We can see that the electron bunches
in the 3-rd wake period have finite transverse momentum beamlets (indicated
by the arrows) and have the y-component of their momentum equal to
approximately 4. This corresponds to the generic property of the wakefield
acceleration, when the electron is trapped both in the longitudinal and
transverse directions. According to Ref.\cite{T-D} the characteristic
timescale of the electron longitudinal motion is equal to the acceleration
time, $t_{acc}=\omega_{pe}^{-1} \gamma_{ph}^2$. Here 
$\gamma_{ph}=\omega/ \omega_{pe}$ is the gamma factor associated with the wake wave phase
velocity, $v_{ph}$, i.e. $\gamma_{ph}=1/\sqrt{1-(v_{ph}/c)^2}$ (see for
details review article \cite{MTB} and quoted literature). In the transverse
direction, the electrons being injected with a finite transverse momentum
(see Refs. \cite{TWB, CH}), move along a wiggling trajectory, also called
betatron oscillations, in which the frequency and amplitude change, with a
typical timescale given by $t_b=\omega_{pe}^{-1} \gamma_e^{1/2}$ \cite{BETA}, 
where $\gamma_e \approx \gamma_{ph}^2$ is the fast electron relativistic
gamma factor. We point out that the beam transverse dynamics determines its
transverse emittance (e.g. see Ref. \cite{CPT}). From the 2D-PIC simulation
results, presented in Fig. 5, the electron beam divergence is estimated to
be in the range from 10 to 20 mrad, which is in sufficiently good agreement
with the experimental data as seen in Fig. 2 (b).

\section{Summary}

In conclusion, when a 3 TW laser beam with a pulse duration equal to 70 fs
is focused by a f/13 Off-Axis-Parabolic mirror into a He gas jet target of
length 600 $\mu m$, the quasi-monoenergetic electron electron beams are
observed. By changing the target plasma density the fast electron energy and
their energy spectrum change in a controllable way. The 2D PIC simulations
demonstrate that the wake field generation occurs in the self-modulated
laser wake field acceleration regime when multiple fast electron bunch
generation occurs and the bunches are well localized both in energy and
coordinate space. The PIC simulations qualitatively reproduce the dependence 
of the experimentally observed energy spectrum on the plasma density.

\section*{Acknowledgements}

We acknowledge Prof. T. Tajima for fruitful discussions and Dr. T. Zh.
Esirkepov for providing the 2D-PIC simulation code. This work was partly
supported by the Ministry of Education, Culture, Sports, Science and
Technology of Japan, Grant-Aid for Specially Promoted Research No. 15002013.

\newpage

\section*{Figure captions}

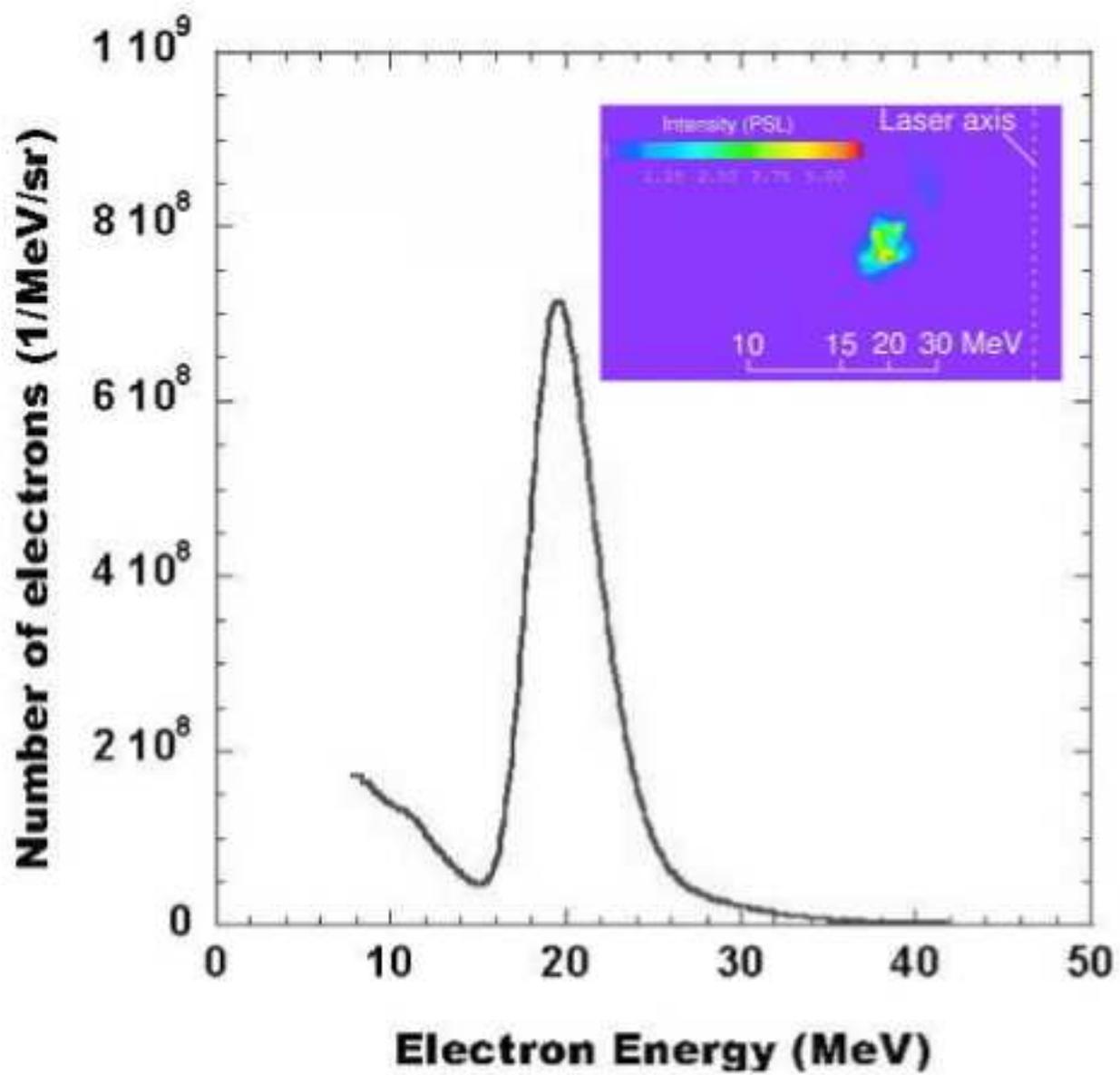
\noindent Fig. 1. Typical energy distributions of the electron beam and image of the energy resolved electron beam. 
A quasi-monoenergetic electron beam with an energy of 19.6 MeV observed at a plasma density equal to $5 \times 10^{19}$ cm$^{-3}$.

\bigskip

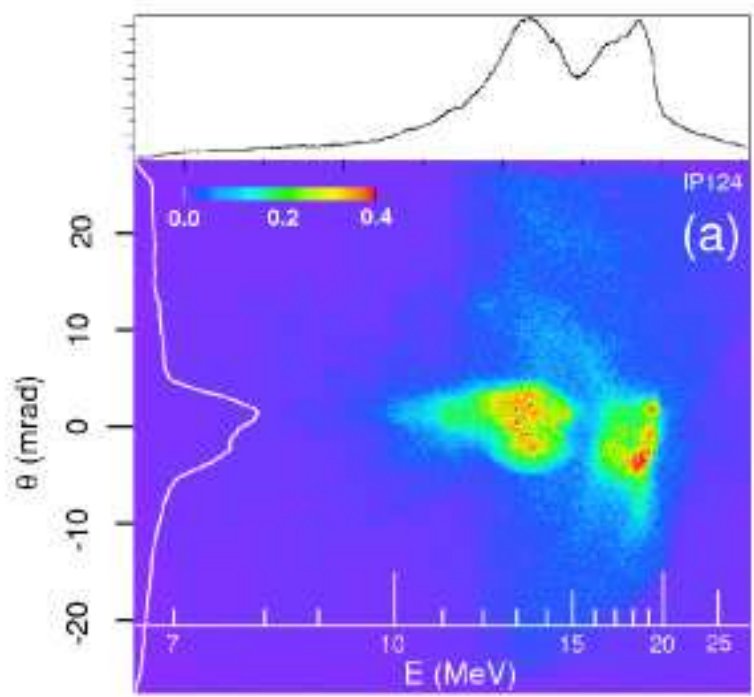
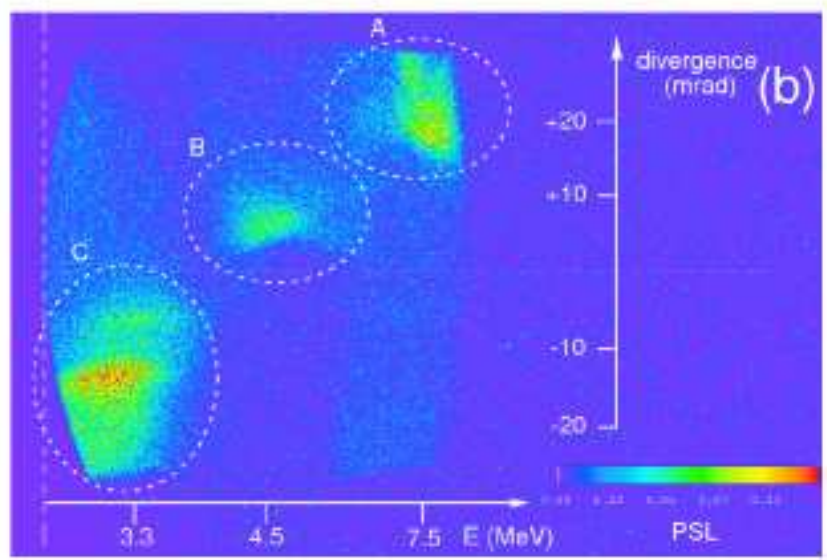
\noindent Fig. 2. Images experimentally obtained 
in a single shot of the spatial (along the vertical axis) and energy (along the horizontal axis) electron distribution
with: 
(a) a single spatial peak and double energy peak; 
(b) triple spatial  peak and triple energy peak. 

\bigskip

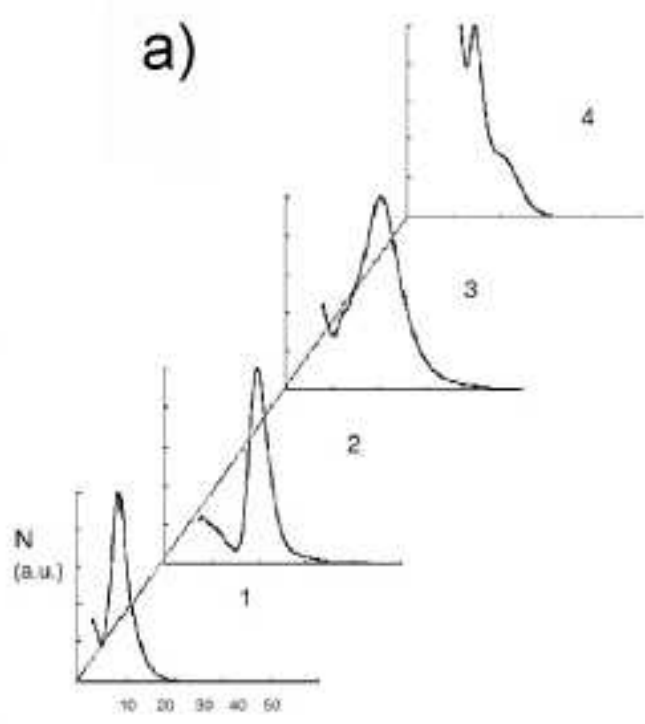
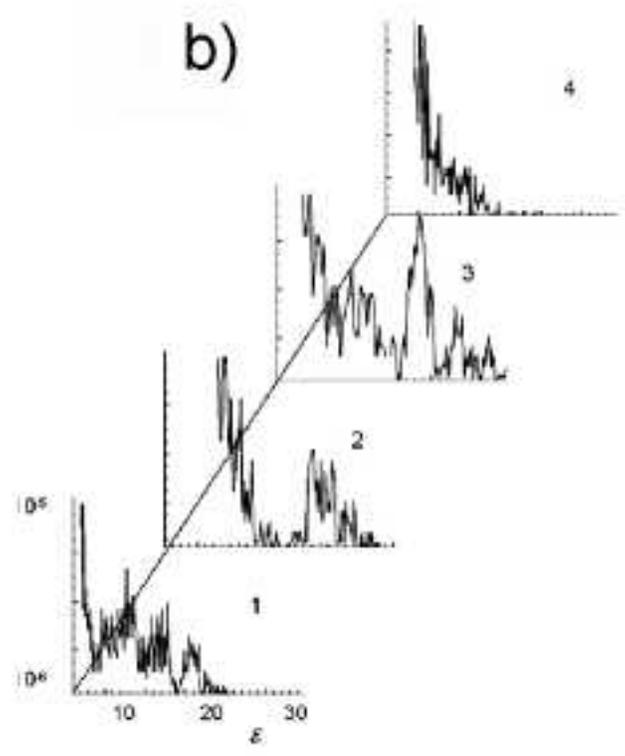
\noindent Fig. 3. Electron energy spectrum versus plasma density. 
a) experimental data and b) simulation results for
$n_e=4.1\times 10^{19}cm^{-3}$ (1); 
$n_e=4.7\times 10^{19}cm^{-3}$ (2); 
$n_e=5\times 10^{19}cm^{-3}$ (3); 
$n_e=6.6\times 10^{19}cm^{-3}$ (4).

\bigskip

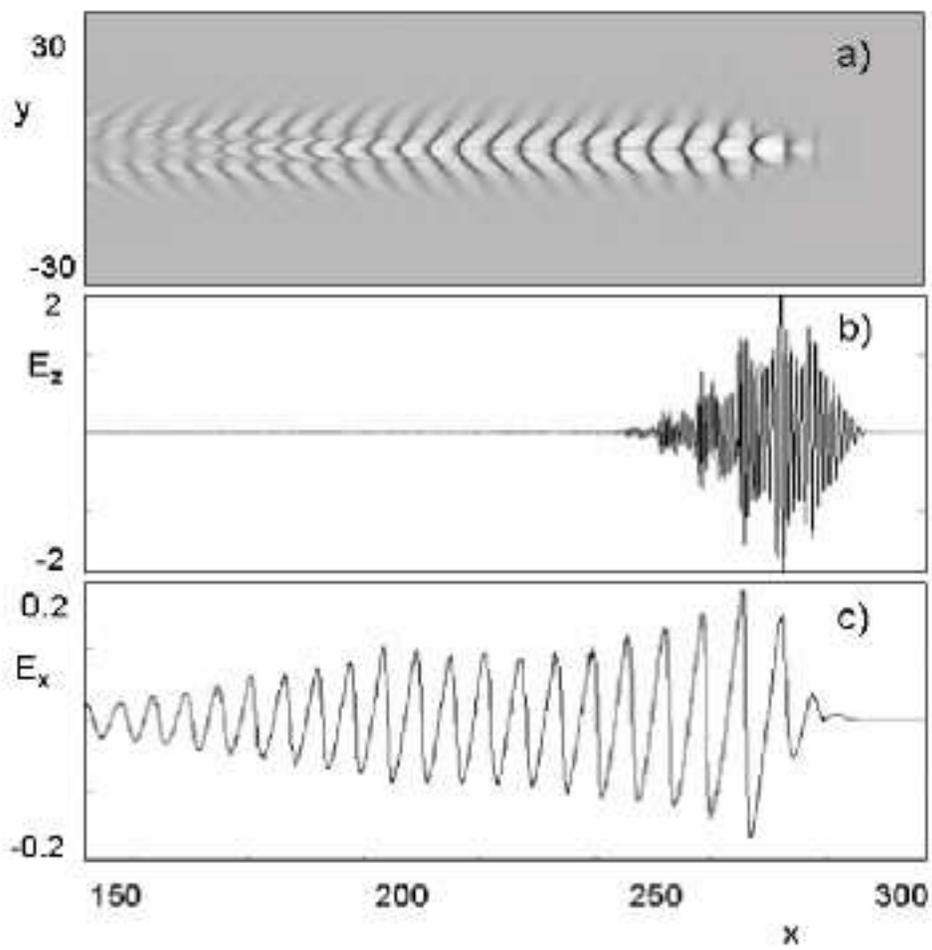
\noindent Fig. 4. Wake wave seen in the electron density, $n(x,y)$, in the $(x,y)$ plane a);
$E_z $ component of the electric field shows the laser pulse modulation b) 
and  $E_x$ component of the electric field of the wake wave, c), along the laser pulse axis, $y=0$, for $t=300$. 

\bigskip

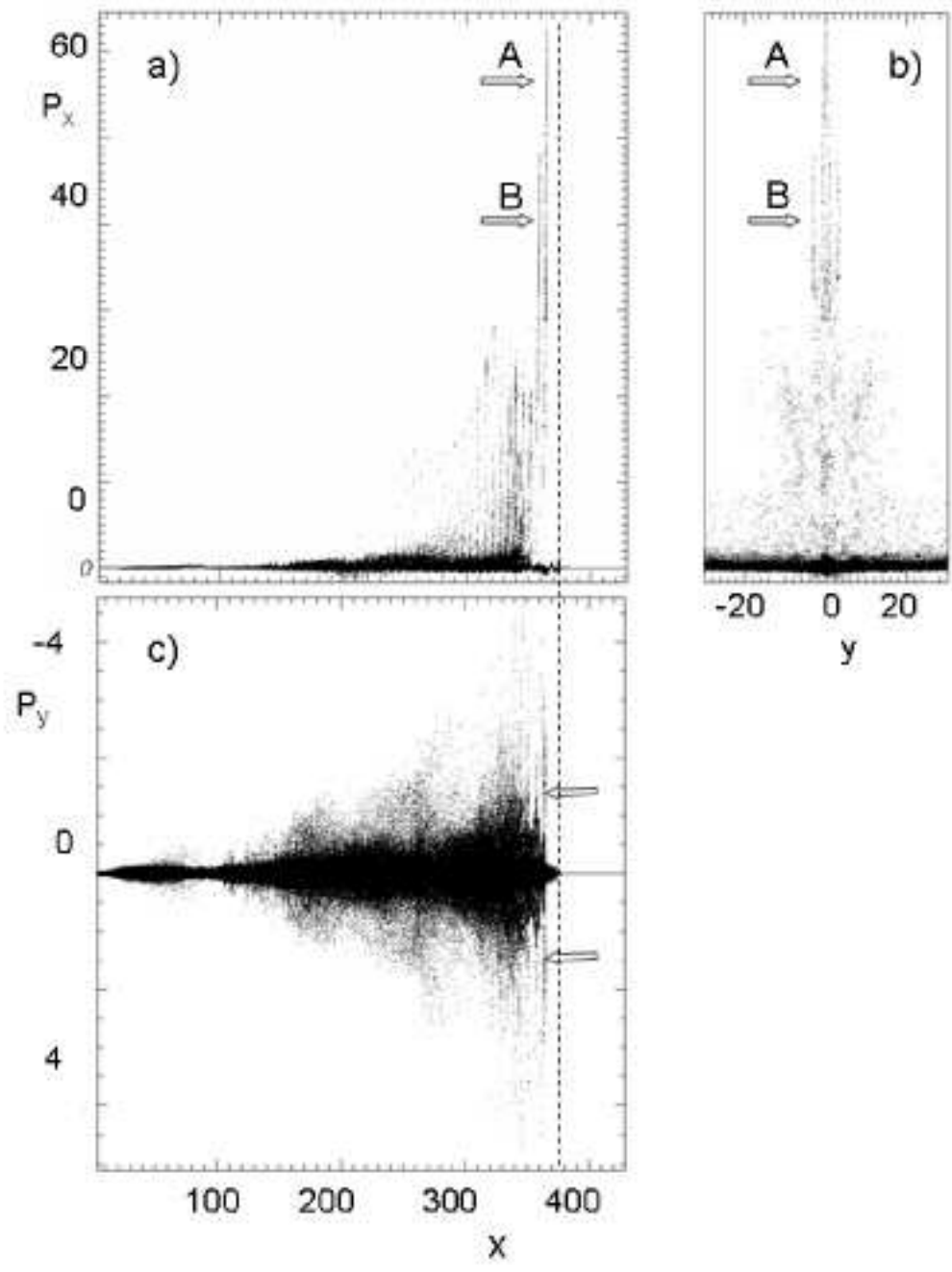
\noindent Fig. 5. Electron phase space: a) plane $(x,p_x)$, b) plane $(y,p_x)$, and c) plane $(x,p_y)$ at $t=400$. 
Arrows indicate the electron beamlets. Dashed line indicates a position of the laser pulse front edge. 

\end{document}